\documentclass[12pt]{article}
\usepackage[dvipsnames]{xcolor}
\usepackage{epsfig}
\usepackage{pstricks, amscd}
\usepackage{cite}
\usepackage{graphicx}
\usepackage{makeidx}
\usepackage{amsmath}
\usepackage{amssymb}
\usepackage{nccmath}
\usepackage{setspace}
\usepackage{ulem}
\usepackage{tensor}
\usepackage{bm}
\usepackage[format=hang, justification=justified, font=small, labelsep=period, figurewithin=section]{caption}

\usepackage[unicode]{hyperref}
\hypersetup{hypertexnames=false, colorlinks=true, linkcolor=BlueViolet, citecolor=ForestGreen, urlcolor=blue, linktoc=all}

\renewcommand\rho{\varrho}
\newcommand{\bi}{\bibitem}
\newcommand{\be}{\begin{eqnarray}}
\newcommand{\ee}{\end{eqnarray}}
\newcommand{\rar}{\rightarrow}

\numberwithin{equation}{section}

\begin{document}   

\begin{titlepage}
%\title{Cosmic rays from heavy particle annihilation}
\title{ Cosmic rays from  annihilation of heavy dark matter particles}
%\author{A.\,D. Dolgov$^{a, b}$ and A.\,S. Rudenko$^{a, c}$}
\author{E.V. Arbuzova$^{a,b}$
A.\,D. Dolgov$^{a, c}$, A.A. Nikitenko$^{c,d}$}

\maketitle
\begin{center}
$^a${Department of Physics, Novosibirsk State University, \\ 
Pirogova st.\,2, Novosibirsk, 630090 Russia} \\
$^b${Department of Higher Mathematics, Dubna State University, 141982 Dubna, Russia}\\
$^c${Bogolyubov Laboratory of Theoretical Physics, Joint Institute for Nuclear Research, \\
Joliot-Curie st.\,6, Dubna, Moscow region, 141980 Russia} \\
$^d${Department of Fundamental Problems of Microworld Physics, Dubna State University, Dubna, 141982, Russia}
\end{center}

%\title{Problems of spontaneous baryogenesis}

%\title{Cosmic rays from heavy particle annihilation}

\begin{abstract}

The origin of the ultra high energy cosmic rays via annihilation of heavy stable, fermions "f", of the 
cosmological dark matter  (DM) is studied. The particles in question are  supposed to be created 
by the scalaron decays  in $R^2$ modified gravity. Novel part of our approach is the assumption
that the mass of these carriers of DM is slightly below than a half of the scalaron mass. In such a case 
the phase space volume becomes tiny. It leads to sufficiently low probability of "f" production, so their average
cosmological energy density could be  equal to the observed energy density of dark matter. Several regions of the universe, where the annihilation could take place, are studied. They include the whole universe under assumption 
of homogeneous energy density, the high density DM clump in the galactic centre, the cloud of DM in the Galaxy with realistic density distribution, and high density clusters of DM in the Galaxy. 
Possible resonance annihilation of $f \bar f$
into energetic light particle is considered.
We have shown that the proposed scenario can successfully explain the origin of the ultrahigh energy flux of cosmic
rays where the canonical astropysical mechanisms are not operative.

\end{abstract}
 
\end{titlepage}

%\maketitle

\section{Introduction}

The origin of the extremely {energetic} cosmic rays with energies exceeding $10^{20}$ eV  remains uncertain 
despite impressive theoretical activity in this field. 
{The} commonly accepted astrophysical mechanisms of ultra 
high energy cosmic {ray} (UHECR)
production through supernova explosion or catastrophic processes in active galactic nuclei encounter serious problems, {that stimulated} alternative 
suggestions of UHECR production by heavy particle decays \cite{Berezinsky:1997hy,Kuzmin:1997jua,Birkel:1998nx} or 
annihilation \cite{Blasi:2001hr,Blasi:2001rb,Dick:2002kp}, see also \cite{Dick:2005mk} for a review. 

As is shown in Ref. \cite{Blasi:2001hr}, to produce the observed flux of cosmic rays,  the cross section of dark matter (DM) annihilation should be:
\be
\langle \sigma v \rangle \sim 10^{-26} \text{cm}^2 (M_X/10^{12}\, \rm{GeV})^{3/2}, 
\label{crsec_Bl}
\ee
where $M_X$ is the mass of dark matter particle.  
The magnitude of the cross-section is extremely high and demands contribution {of partial wave with}
huge angular momenta due to unitarity constraints.  

In this paper only the contribution of the DM particle decays into the highest energy cosmic rays is
considered,  in the range that cannot be explained by usually assumed catastrophic astropysical processes.

{We assume that the superheavy particles
of DM are produced through the scalaron decay in $R^2$ modified gravity~\cite{Starobinsky:1980te}.
The calculations of the scalaron decay probabilities were performed in our paper~\cite{Arbuzova:2021oqa},
as well as in several other ones. In all these works it was supposed that the masses of the decay products were
much smaller than the scalaron mass, $M_R$, which according to Ref. \cite{Faulkner:2006ub} is equal to:}
 \be 
 M_R = 3\times10^{13} \ \text{GeV}.
 \label{M-R}
 \ee

 In our papers \cite{Arbuzova:2018apk,Arbuzova:2020etv,Arbuzova:2021etq}  
 the  production of superheavy carriers of dark matter 
\ via scalaron decays into particles 
with masses up to $M \lesssim 10^{12}$ GeV were studied, so in all cases  the condition $(2 M/M_R)^2 \ll 1$ was satified.

In our recent paper \cite{Arbuzova:2023dif} we studied a mechanism of possible origin of UHECR by the decay of heavy quasistable particles of
dark matter. It is usually assumed that dark matter particles are absolutely stable. However, it was argued by Zeldovich 
\cite{Zeldovich:1976vq,Zeldovich:1977be} that  all massive presumably stable elementary particles should decay through virtual 
black hole formation. It was shown in \cite{Arbuzova:2023dif} that for certain range of parameter values the life-time of the dark matter particles 
could exceed the universe age  by several orders of magnitude. If this is the case the products of their decays could make a
noticeable contribution to the flux of ultra high energy cosmic rays. 

In the present paper we investigate an alternative source of high energy cosmic rays, namely, annihilation of superheavy dark matter particles. 
We assume that these heavy particles of dark matter are fermions directly produced by the scalaron decays. Previous calculations of 
the scalaron decay probabilities have been performed in the limit of low masses of the decay products, much smaller than the scalaron mass, 
$M_R $.  For instance, the width of the scalaron decay into a fermion-antifermion pair has been found to be:
 \be
  \Gamma_{m_f} = \frac{m_f^2 M_R}{6 M_{Pl}^2},
  \label{Gamma-f-old}
  \ee
 where $m_f$ is the fermion mass, $M_{Pl} = 1.22 \times10^{19}$GeV  is the Planck mass\footnote{ We use all over the paper 
 the natural system of units with speed of light $c=1$, Boltzmann constant $k=1$, and reduced 
Plank constant $\hbar = 1$.

The gravitational coupling constant is $G_N = 1/M_{Pl}^2$, where the Planck mass is 
$M_{Pl} = 1.22 \cdot 10^{ 19}\,\rm{GeV} = 2.17\cdot 10^{-5}\,{\rm g} $.
1 GeV$^{-1} = 0.2\cdot 10^{-13}$ cm, 1sec $ = 3\cdot 10^{10}$ cm, and 1yr = $3.16\cdot 10^{7} $ sec.}.   
 For decay width \eqref{Gamma-f-old} the energy density of heavy fermions would be much larger than the averaged cosmological density of dark matter:
 \be
 \rho_{DM} \approx 1 \ \text{keV/cm}^3.
 \label{rho-DM}
 \ee
 
 The probability of the decay would be strongly suppressed by the phase space factor, $\sqrt{1 - 4 M_f^2/M_R^2}$,  
 if the fermion mass, $M_f$, is extremely close to $M_R/2$. 
Fixing the ratio $2M_f/M_R$ sufficiently close to unity, we can adjust the energy density of heavy fermions equal to the observed 
energy density of dark matter \eqref{rho-DM}.

{The phase space factor for the case of a particle (or any other state) with momentum $P$ transforming into
two particles with momenta $p_1$ and $p_2$ is given by the expression:
\be 
\Phi = \int \frac{d^3 p_1\,d^3 p_2}{(2\pi)^6 \,4E_1 E_2} \, (2\pi)^4 \delta^{(3)} \left(\vec{P}-\vec{p_1}-\vec{p_2}\right)
\delta(P_0 -E_1 -E_2).
\label{phi-1}
\ee
It is scalar with respect to Lorenz transformation and can be calculated in any frame. Most convenient is to 
make calculations in the center of mass frame where the space component of $\vec{P}$ is zero, and
due to the delta-function $\vec{p}_1 + \vec{p}_2 = 0$. }
Let us start with the simplest case of relativistic particles for which $E_j = {p}_j$, $j = 1,2$
and $p_{1,2}$ are the absolute values of the particle momenta.

{Using the delta function $\delta^{(3)} (\vec p_1 + \vec p_2) $ we take the integral over $d^3 p_2$ %= 4\pi p_2^2\, dp_2$ 
and obtain
\be
\Phi_{rel} = \int\frac{d^3 p_1 \delta (P_0 - 2 E)}{ (2\pi)^2 \cdot 4 E^2},
\label{phi-2}
\ee
where $E = p_1 $.}

{Since  $d^3 p_1 = 4\pi E^2 dE$  the integral is reduced to
\be 
\frac{1}{4\pi}\,\int dE\, \delta(P_0- 2E) = 1/(8\pi), 
\label{|phi-3}
\ee
the well known result.}

{
Now let us consider nonrelativistic particles, $E = M+ p^2/(2M)$. In this case
\be
\Phi_{nr} = \int \frac{d^3 p \,\delta (P_0 - 2M - p^2/M)}{(2\pi)^2\, 4 E^2} =
\int\frac{dp^2\, p}{8\pi M^2} \,\delta(P_0 - 2M - p^2 /M).
\label{phi-4}
\ee
In our case $P_0 = M_R \approx 2M$, $M=M_f$, and $p^2 = M(P_0- 2M)$ and thus we obtain
%Assuming that $P_0 \approx 2 M$ we obtain 
\be 
\Phi_{nr} = \frac{1}{8 \pi} \sqrt{\frac{P_0 - 2M}{M} } =  \frac{1}{8 \pi} \sqrt{\frac{P_0^2 - 4M^2}{M(P_0+ 2M)} }
=  \frac{1}{8 \pi} \sqrt{1-\frac{4M^2}{M_R^2}}, %\sqrt{\frac{M_R^2}{M(P_0+ 2M)} }
\label{phi-5}
\ee
instead of $1/(8\pi)$.}

Let us present now the expression for the decay products velocity in the center of mass frame.
For nonrelativistic particles $p = vM$, or better $v^2= p^2/M^2$.
From the expression for momentum below Eq.~(\ref{phi-4}) and Eq.~(\ref{phi-5})
we find:
{
\be 
v^2 =\frac{p^2}{M^2} = 1-\frac{4M^2}{M_R^2},
\label{v-2}
\ee
or finally
\be
v=\sqrt{1 -\frac{4M^2}{M_R^2}}.
\label{v-fin}
\ee
}

{There is another way of the production of $f \bar f $ pair by the scalaron field in non-perturbative way, as if fermions are created
by the collective action of $R(t)$. With very weak coupling to  fermions such channel is exponentially suppressed. 
}

{
When we consider annihilation  of a pair of heavy fermion-antifermion into light particles the cross section is  inversely proportional
to their center of mass velocity but the evolution of the fermion density depends upon the product $\sigma v$ and the smallness
of the velocities of the annihilation particles does not have any impact on the result. In other words we should not multiply the 
efficiency of the annihilation by the tiny factor $\sqrt{ 1 - 4M^2/M_R^2}$. %The old result should be valid.
}
 
{ To estimate the necessary value of the phase space suppression factor, we proceed as follows. The scalaron decays not only to heavy 
 fermions but to all other  particles coupled to curvature $R$. { Let us note, that scalaron is not canonically normalised scalar field, 
 as is dicussed e.g. in Ref.~\cite{Arbuzova:2021oqa}. By this reason the factor $1/(8 \pi)$ is absent in the expressions for the decay widths. }   
  In particular according Ref.~\cite{Arbuzova:2021etq}, Eq. (142)
 the decay width of of $R$ to massless scalars is equal to:
 \be
 \Gamma_S = \frac{M_R^3 N_S}{{24} M_{Pl}^2},
 \ee
  where $N_s$ is the number of massless or light scalar species. A 
 reasonable guess is $N_s \sim 100$.
The energy density of the produced scalars, Eq. (161) of Ref.~\cite{Arbuzova:2021etq},  is:
\be
\rho_S = \frac{M_R^3 N_S}{{240} \pi t}.
\label{rho-S}
\ee
The width of the scalaron decay into massive fermions with an account of the phase space suppression factor 
 (\ref{phi-5}) can be read-off 
Eq. (187) of Ref.~\cite{Arbuzova:2021etq} and is equal to:
\be
\Gamma_f = \frac{M_R M_f^2}{{ 6} M_{Pl}^2} \sqrt{1 - \frac{ 4 M_f^2}{M_R^2}}. 
\label{Gamma-f-1}
\ee}

The transition from the R-dominated regime to GR takes place at the moment of the complete
decay of the scalaron. We assume that the decay into massless scalars  is the dominant one. 
So we must take $1/t_{GR} = \Gamma_S$ Hence the energy density of scalars at the moment to transition to GR is
\be
\rho_S { (t_{GR})} = \frac{M_R^6 N_S^{2}}{24\cdot 240 \pi M_{Pl}^2},
\label{rho-s-GR}
\ee%%Let us denote the phase space factorÃÂ  $ \sqrt{1 - 4 M_f^2/M_R^2} $ as $P$.
The energy density of heavy fermions at the same moment $ t_{GR}$ of transition to GR { (see also Eq. (190) in Ref.~\cite{Arbuzova:2021etq}):} 
\be
{\rho_f (t_{GR}) = \frac{M_R M_f^{ 2}}{120 \pi t_{GR}}\, \sqrt{1-\frac{4M^2}{M_R^2}} .}
\label{rho_f-GR}
\ee
The energy density at the beginning of GR is dominated by massless scalars and the temperature at the moment of transition 
to GR is determined by the equation:
\be
\rho_S = \frac{M_R^6 N_S^{ 2}}{24\cdot 240 \pi M_{Pl}^2}  = \frac{\pi^2}{30} \,g_* T_{GR}^4.
\label{rho-S-GR}
\ee
Correspondingly the temperature of the universe at this moment  was
\be
T_{GR}^4 = M_R^4 \left(\frac{N_S^{ 2}}{192 \pi^3 g_*}\right)  \left(\frac{M_R}{M_{Pl}}\right)^2.
\ee
So $T_{GR} \approx {6} \cdot 10^{-4} M_R $.
We know that today the energy density of DM is about {  $4 \cdot 10^{3}$ } of the energy density of {the microwave photons,} i.e. 
$\rho_f/\rho_S \approx 4 \cdot 10^{3}$, {(since the energy density of the CMB photons is 0.26 eV/cm$^3$)}.
The redshift of  the temperature $T_{GR}$ {to the present day CMB temperature $T_{CMB} = 2.7 K = 2.35 \cdot 10^{-4}$ eV
is:
\be
 \frac{ T_{GR}}{T_{CMB}} = \frac{{ 6 \cdot } 10^{-4} M_R} { 2.7K } { \approx 8} \cdot 10^{22}. 
 \label{z-CMB}
 \ee
  In fact this ratio should be somewhat smaller due to contribution of three light
neutrinos to relativistic matter density. So for future estimates we take this ratio as $10^{22}$.}
Since the energy density of relativistic matter drops down as $1/a^4$ and of nonrelativistic matter drops as $1/a^3$, the present day ratio of 
the densities of the heavy fermions making dark matter to the relativistic matter density  
would increase by the factor 
$z  \sim 10^{22}$. {From Eqs. \eqref{rho_f-GR} and \eqref{rho-S} follows that the ratio of the densities of nonrelativistic dark matter
to the total cosmological energy density of relativistic particles would be:}
\be {
\frac{\rho_f (t_{GR})}{\rho_S {( t_{GR})}} = 
\frac{M_R M_f^{2}}{120 \pi t_{GR}}\cdot \frac{ 240 \pi t_{GR}}{{M_R^3 N_S}} \sqrt{1 - \frac{ 4 M_f^2}{M_R^2}}  =  
5 \cdot 10^{-3} \sqrt{1 - \frac{ 4 M_f^2}{M_R^2}} , }
\label{ratio-of-rho}
\ee 
where we took $M_f = M_R/2$ and $N_S =100$.
With the enhancement  factor $z=10^{22}$ this ratio at the present time 
would be $5 \cdot 10^{19}\sqrt{1 - 4 M_f^2/M_R^2}$. 

As we mentioned above the ratio of the DM energy density to the energy density of relativistic matter today is 
$\rho_f/\rho_S \approx 4 \cdot 10^{3}$.
So to make the proper density of DM equal to 1 keV/cm$^3$ we need the phase space suppression to be about $10^{-16}$ . This is unnaturally tiny.

{However, making this conclusion we relied on adiabatic universe expansion but this is certainly not the case.
There are several known phase transitions (PT), possibly first order ones with strong overcooling
leading to some quasi-inflationary stages. 
For example the QCD phase transition, from the quark confinement phase to the hadron phase is probably the first order one.
It may significantly diminish the ratio of dark matter density with respect to the relativistic one. However,  it would also diminish
cosmological baryon asymmetry since most probably it was generated prior to QCD PT. }

{First order electroweak phase transition (EW PT) can lead to significant supercooling, as is studied in many papers, see e.g. ~\cite{Gould:2022ran}
and references therein. As is argued in Ref.~\cite{Cline:2012hg} particle physics models which account for dark matter or which lead to successful 
baryogenesis may predict a strongly first-order electroweak phase transition.
Some discussion of first order phase transitions with supercooling can be found in Ref.~\cite{Cline:2018fuq}.
There can be quite a few  first order phase transitions at high temperatures, so the necessary enhancement  factor could be strongly diminished.
In a sense it is a free parameter of the theory.
One should take care of preserving baryon asymmetry, so these phase transition should be prior to asymmetry generation, but it can be done.
So to conclude we may increase the phase space suppression factor to a reasonable value so the considered below
resonance annihilation would not suffer.
} 

The paper is organised as follows. In the following section  the cosmological energy density of superheavy
fermions is calculated in the case when their mass is very close to a half of the scalaron mass.
In section \ref{flux-CR}  the contribution of the $f \bar f$ - annihilation into the  flux of UHECR is estimated for different
spatial distribution of f-particles: homogeneous cosmological one,  high density clump in the galactic centre,
and finally we conclude.

\section{Energy density of heavy fermions in the universe \label{ener-dens} }

The ratio of number density of heavy fermions (in what follows we will call them $f$-particles), $n_f$, to the total number density, $n_{tot}$, of relativistic plasma created by the scalaron to the moment of its complete decay can be estimated as:
\be
\frac{n_f}{n_{tot}} = \frac{\Gamma_f}{\Gamma_{tot}}, 
\label{number-ratio} 
\ee
where $\Gamma_f$ is given by Eq. \eqref{Gamma-f-1} and $\Gamma_{tot}$ is the total width of the scalaron decay. We assume that the scalaron 
predominantly decays into massless scalar bosons minimally coupled to gravity. In this case its width is equal to:
\be
\Gamma_{tot} \approx \Gamma_s = \frac{M_R^3}{24M_{Pl}^2}.
\label{gamma_s}
\ee    
Thus, we obtain:
\be
\frac{n_f}{n_{tot}} = \left(\frac{2M_f}{M_R}\right)^2\sqrt{1- \frac{4M_f^2}{M_R^2}} 
\approx \sqrt{1- \frac{4M_f^2}{M_R^2}}, 
\ee
since $M_f \approx M_R/2$.

The energy density of the heavy fermions can be presented in the from:
\be
\rho_f = M_f \,n_f = M_f\, \frac{n_f}{n_{\gamma}}\, n_{\gamma}, 
\label{rho-f}
\ee
where $n_{\gamma} = 412$ cm$^{-3}$ and the ratio $n_f/n_{\gamma}$ slowly changes in the course of the universe evolution. 
According to our assumption $M_f \approx M_R/2 =  1.5 \times 10^{13}$~GeV, see Eq.~\eqref{M-R}.
To achieve the value of the fermion energy density equal to the observed cosmological DM energy density 
in the present day universe \eqref{rho-DM},  
$\rho_f \approx \rho_{DM}$, we must take:
\be
 \frac{n_f}{n_{\gamma}} = 1.6 \times 10^{-22}. 
 \label{n-f-over-n-gam}
 \ee
The ratio of $n_f/n_{\gamma}$ in the early universe depends upon the law of the cosmological expansion and the spectrum 
of elementary particles. Slightly varying the difference $(M_R - 2M_f)$ one can obtain the initial value 
of $n_f/n_{\gamma}$ that could provide the necessary final density  of DM \eqref{n-f-over-n-gam}.
 This density could be much smaller than the  frozen density of $f$-particles, 
if one starts from their equilibrium  density. Indeed, 
when calculating the asymptotic density of stable $f$-particles,  normally the Zeldovich equation is used under assumption of an initial equilibrium value of their density. The decrease in the density of $f$-particles occurs due to their mutual annihilation. The result does not depend on the initial value of $n_f$ and is equal to 
$ n_f  = n_\gamma /(\sigma_{ann} m_f M_{Pl} )$ up to a logarithmic factor. But if the initial value of $ n_f$ is less than that given above, then their annihilation is  insignificant and $ n_f$ remains practically equal to its small initial value.
 
%\tcmag{\section{Comments on system of units \label{s-units} - {\red fotenote on p.2 }}}

%We use all over the paper the natural system of units with speed of light $c=1$, Boltzmann constant $k=1$, and reduced 
%Plank constant $\hbar = 1$.

%The gravitational coupling constant is $G_N = 1/M_{Pl}^2$, where the Planck mass is 
%$M_{Pl} = 1.22 \cdot 10^{ 19}\,\rm{GeV} = 2.17\cdot 10^{-5}\,{\rm g} $.
%1 GeV$^{-1} = 0.2\cdot 10^{-13}$ cm, 1sec $ = 3\cdot 10^{10}$ cm, and 1yr = $3.16\cdot 10^{7} $ sec.

%---------------------------------------------------

 \section{Flux of cosmic rays from  $f$-particle annihilation\label{flux-CR}} 
 
 \subsection{Energy spectrum of the annihilation products \label{flux-annih}} 

The flux of high energy particles is determined by the cross-section of annihilation of heavy fermions.
We use the natural estimate for the annihilation cross section, namely:  
\be
\sigma_{ann} v \sim \alpha^2 \,g_*/M_f^2,
\label{cr-sec} 
\ee
where $v$ is the centre-of-mass velocity,  
  $\alpha$ is the coupling constant with the typical value $\alpha \sim 10^{-2}$, and $g_*$ is the number of the open annihilation 
  channels,  $g_* \sim 100$. 
  With $M_f = 1.5 \cdot 10^{13}$ GeV we estimate $\sigma_{ann} v \sim 2 \cdot 10^{-56} \text{cm}^2$. This value is smaller than the  cross 
  section \eqref{crsec_Bl} at least by 30 orders of magnitude. 
{To smooth down this problem we have suggested a way to enhance the efficiency of the annihilation. }
  
The rate of the decrease of the $f$-particle density per unit time and volume is equal to:
\be
\dot n_f = \sigma_{ann} v n_f^2 = \alpha^2 g_* n_f^2/M_f^2, 
\label{dot-n-f}
\ee

{We assume, though it is not necessary, that the annihilation is sufficiently slow, so that the number density $n_f$ does not essentially change
during the universe age.}

The annihilation of heavy $f$-particles leads to a continuous contribution to the rate of cosmic ray production 
% process leads to the contribution of annihilation of heavy f-particles to the rate of the energy density of the created cosmic rays 
per unit time and unit volume equal to:
\be 
\dot \rho_f = 2M_f \dot n_f .
\label{dot-rho-f3}
\ee
{The results \eqref{dot-n-f} and \eqref{dot-rho-f3} are valid for the total flux integrated over particle energy.}

{To compare our results with observational data we need to know the energy distribution of  the 
cosmic ray particles produced in the process of $f \bar f$ - annihilation. We postulate that the differential energy 
spectrum of the number density flux, $\dot n_{PP} (E)$, of the produced particles (PP)  is stationary and has the form, which we think is 
 reasonable:}
\be 
\frac{d \dot n_{PP} (E)}{dE}  =  {\mu^3}\, \exp \left[ -\frac{(E - { 2} M_f/\bar n)^2}{\delta^2}\right] \theta(2M_f-E) .
\label{dn-dE}  
\ee
{Here $\mu$ is a normalisation factor, with dimension of mass or, what is the same, of inverse length,}
 to be determined in what follows, $\bar n$ is the average number of particles created in the  
process of $f\bar f$ - annihilation. % and $m_0$ is some mass relevant to the scale of the energy density of the ultrahigh energy cosmic rays. 
%We took $ m_0 = 1$ eV.
This distribution ensures maximum energy of the annihilation products 
$E_{max} = 2M_f$ and { the average energy per one particle is equal to
$\bar E \approx 2M_f/\bar n$}, if the width of the distribution, $\delta$, is  sufficiently small. 
The value of $\delta$ will be adjusted to the observed spectrum of the cosmic rays, see below.

The contribution from the heavy particle annihilation into cosmic ray flux is equal to:
\be 
\frac{d \dot \rho_{PP} (E)}{dE} = E\, \frac{d \dot n_{PP} (E)}{dE}.
\label{drho-de} 
\ee

Correspondingly the total flux of the energy density of the produced particles with  the number density 
spectrum \eqref{dn-dE} is: 
\be
%\frac{ \rho_{PP} (E)}{dE} 
\dot \rho_{PP} =  \int_0^{2M_f}  E  \left(\frac{d\dot n_{PP}(E)}{dE}\right)\,dE\, .
\label{dot-rho-f-2}
\ee
We assume, that $\dot \rho_{PP} = const$, since the observed flux of the cosmic rays is stationary.

In the case of $\delta \ll M_f$, that is assumed to be true, we find neglecting terms of the order of $ \delta^2$:
\be
\dot \rho_{PP}  = (\sqrt{\pi}/2) {\mu^3}\,\bar M  \delta \, \left[ 1+ Erf( z) \right] \approx \sqrt{\pi}\, {\mu^3}\,\bar M  \delta \,,
\label{int-d-rho}
\ee
where $\bar M = { 2} M_f /\bar n$ and  $z = (2M_f -\bar M)/\delta $, $z \gg 1$ for  small $\delta $.  The error function is defined as:
\be 
Erf( z) = \frac{2}{\sqrt \pi} \int_0^z e^{-x^2} \,dx 
\label{erf} 
\ee
and the Euler-Poisson integral is:
\be
\int_{-\infty}^{+\infty} e^{-x^2} \,dx = \sqrt{\pi}.
\ee
\\

\begin{figure}[htbp]
		%\vspace{-0.5cm}
		\begin{center}
		\includegraphics[scale=0.5,angle=-90]{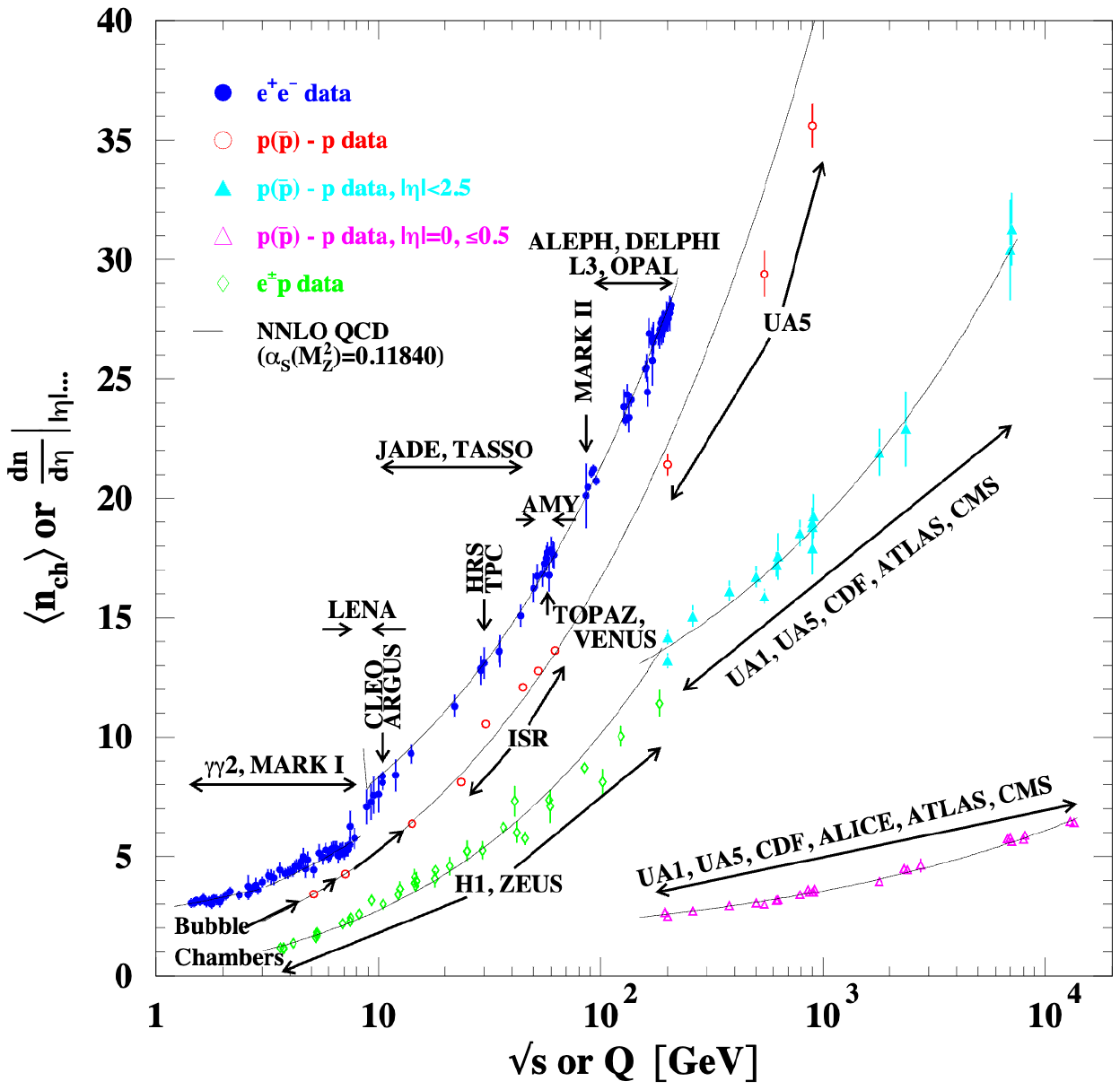}
		\vspace{-3mm}
       \end{center}
	\caption{Average charged-particle multiplicity $<n_{ch}>$ as a function of $\sqrt{s}$
	or $Q$  for   $e^+e^-$ and  $\bar p p$ - annihilations and $pp$ and $ep$ collisions, see Ref.~\cite{PDG-21}. } %\\[2mm]
		\label{f-multip}
	\end{figure}

{
According to the results presented in Refs.~\cite{Cafarella:2006ck,ALICE:2015olq,dEnterria:2019csw} particle multiplicity in high energy proton-proton or
proton-antiproton collisions rises linearly with energy of the colliding particles:
$\bar n \sim 0.01\sqrt{s} $, where $s$ is the square of the center of mass energy of the annihilating particles,
see e.g. Fig.~19.6 of Particle Data Group~\cite{PDG-21}, section 
''Fragmentation functions in e+ e-, ep, and pp collisions''
 at $\sqrt{s} = 10^3 $ GeV and the QCD coupling constant $\alpha_s \approx 0.1$.
We reproduce Fig.~19.6 from ~\cite{PDG-21}  in  Fig.~\ref{f-multip}.}
 
If this law could be naively extrapolated up to $\sqrt{s} \sim 10^{13} $ GeV, as is our case, the average  number of the created
particles could be huge, as large as $10^{11}$.  However, at extremely high energies the particle interaction strength is expected to
drop down significantly and the rate of particle "multiplication" could be by far smaller. The standard GUT predicts $\alpha_s \sim 0.01$ 
at $E \sim 10^{13}$ GeV  but it is not excluded that $\alpha_s$ is even smaller. The probability of multiparticle production is suppressed 
as $\alpha_s ^{\bar n}$, but enhanced by the increasing phase space. 
For the sake of estimate we assume that $\bar n \sim 10^3$, though it is not particularly important to us.

The normalisation factor ${\mu^3}$ in equation (\ref{dn-dE}) can be found from the condition of equality of the total
energy density of the produced cosmic ray particles  per unit time, to the rate of the energy density, $\dot \rho_f$, emitted in the 
process of the  $\bar f f$ - annihilation as given by Eq. \eqref{dot-rho-f3},
 %\be
%\dot \rho_f = 2M_f \dot n_f,
%\label{dot-rho-f6}
%\ee
where $\dot n_f$ is determined by Eq.~(\ref{dot-n-f}). By assumption $n_f$ remains almost constant and a small fraction of 
annihilating $f \bar f$ supplies the high energy cosmic rays flux.

The fraction of $f$-fermion contribution to the mass density of dark matter is equal to $r_{DM} $, implying that
the cosmological density of $f$ is equal to $\rho_{DM} = r_{DM}(\rm{keV/cm})^3$. In what  follows we assume that $r_{DM} = 1$,
so all dark matter consists of f-fermions. 
Taking $n_f = \rho_{DM} / {2 M_f}$ and $M_f = 1.5\times 10^{13} $ GeV we find for the total energy rate of cosmic rays
created by $f\bar f$ -annihilation:
\be
\dot\rho_f^{(ann)} = { 1.48}\cdot 10^{-54} \,\text{GeV}^{-1} \text{cm}^{-6}.
\label{dot-rho-f-ann}
\ee

The constant ${\mu^3}$ can be calculated from the condition of equality of  $\dot\rho_{PP}$ from Eq. (\ref{int-d-rho}) and 
$\dot\rho_f^{(ann)}$ from Eq.~(\ref{dot-rho-f-ann}):  
\be
{\mu^3} = \frac{1.48\cdot 10^{-54} \bar n}{2\sqrt{\pi} {\rm GeV \cdot cm^6} M_f \delta} =
{\frac{2.2 \cdot 10^{-109}\, \bar n}{{\rm cm^3}} \,\left(\frac{\rm GeV}{\delta}\right).}
\label{C}
\ee

{\subsection{Comments on observational data \label{s-obs}}}

\begin{figure}[htbp]
	\vspace{-6.5cm}
	\centering 
  %%%		\begin{minipage}[b]{0.85\textwidth}
  		%\includegraphics[scale=0.40]{fig_UHECR.pdf}
		\includegraphics[width=\textwidth,scale=0.4]{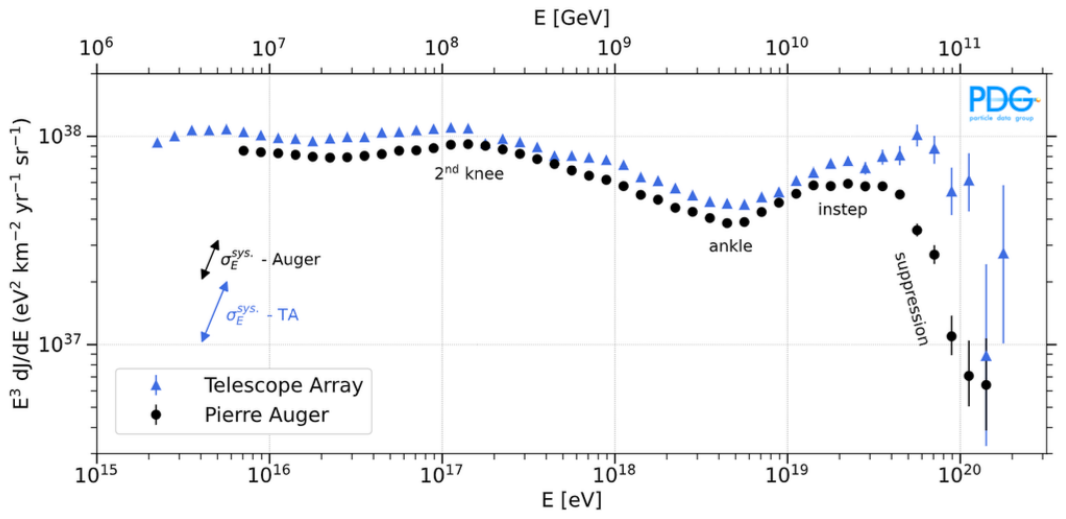}    
%%%		 \end{minipage}
		 \vspace{-8cm}
		\caption{{{The flux of cosmic rays taken from  Fig. 30.7 of Ref.~\cite{PDG-2023}.
	%%%	\bf{It would be better to delete the above figure capture from PDG} 
		}}}
		 \label{f:PDG-spectrum}
\end{figure}

{Theoretically estimated contribution to the observed flux of the cosmic rays should be compared with the data from
the Particle Data Group of 2023~\cite{PDG-2023}. We use  their figure 30.7 (ours Fig.~\ref{f:PDG-spectrum}), where the flux $ dJ/dE$
is presented. Note that the dimension of $dJ/dE$ is:  $[dJ/dE] =  [eV^{-1} km^{-2} yr^{-1}] = [cm^{-2}]$.
The original data on the flux are presented in~\cite{rev-auger}.}

\begin{figure}[htbp]
		\vspace{-0.5cm}
		\begin{center}
		\includegraphics[scale=0.35,angle=-90]{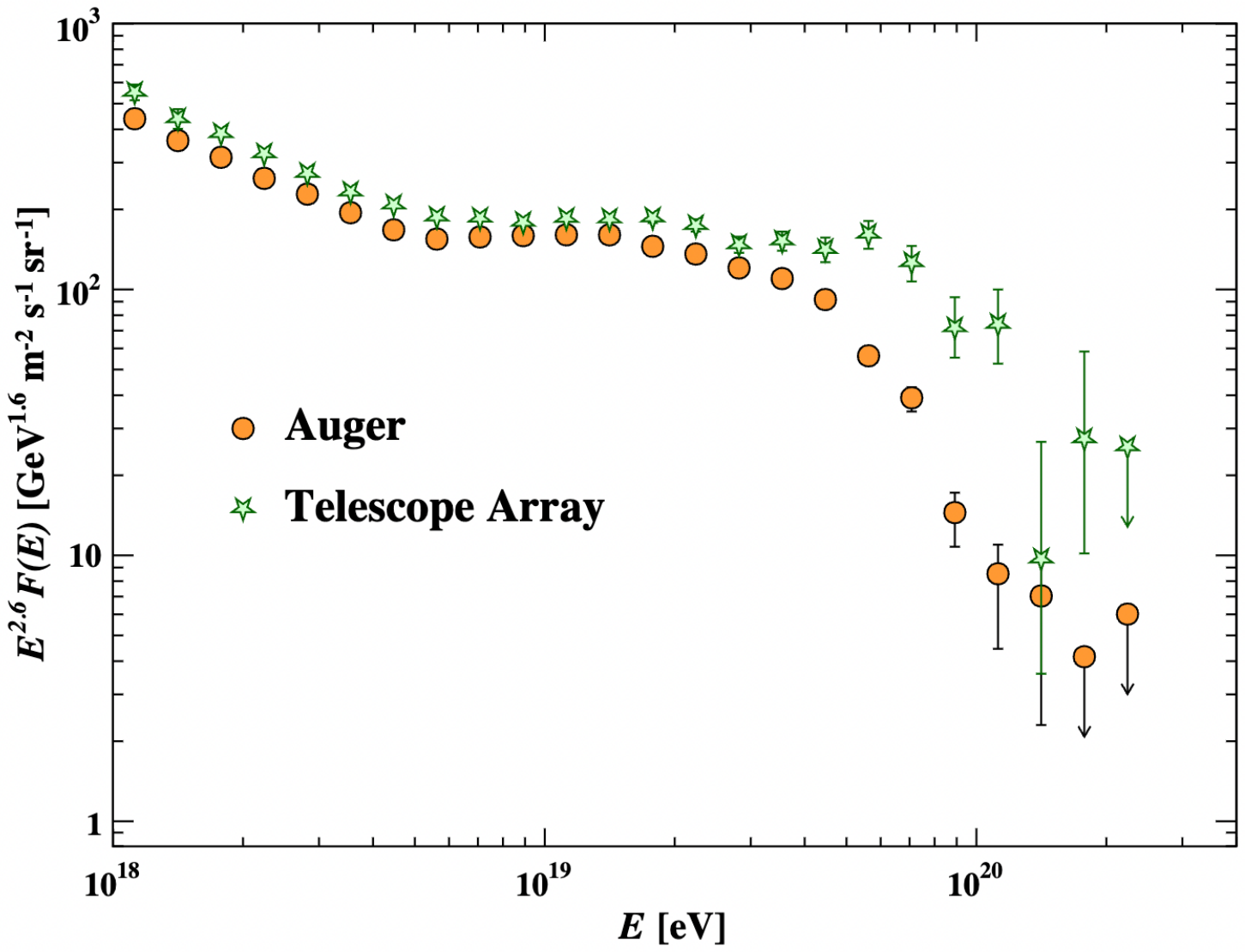}
		\vspace{-3mm}
       \end{center}
	\caption{UHECR flux observations, from Fig. 30.10 of Ref.~\cite{PDG-21}. %\\[2mm]
	}
	\label{fig_UHECR}
\end{figure}

{Similar quantity, denoted as $F(E)$, is presented in Fig.~\ref{fig_UHECR}. This figure is taken from an earlier version of
Particle Data Group of 2021~\cite{PDG-21}, their figure 30.10. The dimension of $F(E)$ is the same as $dJ/dE$,
 $[F(E)] = [GeV^{-1} m^{-2}  s^{-1}] = [cm^{-2}]$, presented in Fig.~\ref{f:PDG-spectrum}.}

{It is instructive to compare the fluxes depicted in Fig.~\ref{f:PDG-spectrum} and Fig.~\ref{fig_UHECR}, say at \\
$E=10^{18} $eV:
\be
\frac{dJ}{dE} = \frac{10^{38}}{10^{54}} \left(eV\cdot km^2 \cdot year \cdot sr \right)^{-1}  = 2\cdot 10^{-49} {\rm cm^{-2}}.
\label{dJ-dE-num}
\ee}
{The flux $F(E)$ at the same energy $E=10^{18} $eV presented in  Fig.~\ref{fig_UHECR} is equal to:
\be
F \approx 5\cdot 10^2 \times 10^{-9\cdot 2.6} \left(GeV\cdot m^2\cdot sec \cdot st \right)^{-1} = 1.3 \cdot 10^{-49}  {\rm cm^{-2}}.
\label{F-num}
\ee}
{The coincidence between these two estimates  is quite good.}\\[2mm]

\subsection{Flux of cosmic rays from homogeneous dark matter \label{s-homDM}}

Let us estimate the energy flux of the products of  the annihilation of dark matter particles 
"in the entire Universe" and reaching Earth's detectors, assuming that dark matter in the Universe is distributed uniformly and isotropically. 

Let us calculate the flux of cosmic rays for the spherical volume of radius $R$ assuming homogeneous distribution of $f$-particles.
We take  $R_{max} \approx10^{28}$cm, since above this distance the redshift cutoff is essential, and finally estimate  the energy  flux 
from the whole universe with (unrealistic) homogeneously distributed dark matter as follows. 
The flux created by source $S$ from the spherical layer with radius  $R$  and width $\Delta R$ would be 
%The differential energy flux of cosmic rays on the Earth created by source $S$  homogeneously distributed 
%originated from the volume element $\Delta V$ at distance $R$ would be:
\be 
\Delta L = \frac{S}{4 \pi R^2}\times 4\pi R^2 \Delta R = S \Delta R.
\label{Delta-F}
\ee
Integrating over the homogeneity scale we find the total flux:
\be
L _{hom}= S R_{max}.
\label{F}
\ee
In the case under consideration 
\be
S_{hom} = \frac{d \dot n_{PP}}{dE}, 
\label{S-of-rho}
\ee
where $(d \dot n_{PP}/dE)$ is given by Eq. (\ref{dn-dE}) with $\mu^3$ determined by expression (\ref{C}).
{Note that dimension of $S$ is $[eV^3] $ or $[cm^{-3}]$, and hence the dimension of $L$ is $[1/cm^2]$. }
%It can be rewritten  as $GeV^{-1}\,cm^{-2}\,sec^{-1}$.
Now we can calculate the contribution to the
flux of high energy cosmic rays, emerging from the $f \bar f$ annihilation, as:
\be 
L _{hom}=  \frac{2.23 \cdot 10^{-109}\cdot 10^{28}\, \bar n}{cm^2}\,\left(\frac{\rm{GeV}}{\delta} \right)
\exp \left[ -\frac{(E - { 2} M_f/\bar n)^2}{\delta^2}\right] \theta(2M_f-E).
\label{L}
\ee

%\tcmag{In the equation above we have taken  $R_{max} \approx10^{28}$cm, since beyond this distance the 
%redshift cutoff is essential.}

{The calculated flux $L$ originated from heavy particle annihilation 
{ should be added either to the flux $dJ/dE$ presented in Fig.~\ref{f:PDG-spectrum} } or to $F(E)$ in Fig.~\ref{fig_UHECR}.
But is should be taken into account that "our" high energy particles contribute to the ultra high energy cosmic rays
in rather narrow range near energies $E \sim 2 M_f /\bar n$, presumably around $E\approx 10^{20}$ eV, where the conventional
astrophysical sources are not efficient. This was the stimulating fact for searching additional possible sources for creation
of UHECR such as heavy particle decays or annihilation.}

A crude order of magnitude estimate of $L$ \eqref{L} assuming $\bar n = 10^3$ and $\delta \sim 1$ GeV would be
$L_{ hom}\sim 10^{-78}$\,cm$^{-2}$. 
We checked that $dJ/dE$ and $F$ are equal to each other at $E= 10^{18} $ eV, taking the values $2\cdot 10^{-49}$.
As one can see  from Fig.~\ref{f:PDG-spectrum},  $dJ/dE$ evolves as inverse energy cube, so at $E \sim 10^{20}$ eV,
but at $E=10^{20}$ eV $E^3 dJ/dE$ remains more or less the same. Hence 
we find $dJ/dE \approx 10^{-55}$\,cm$^{-2}$. As for $F$, we see from Fig.~\ref{fig_UHECR} that $F\times E^{2.6}$
decreases roughly by factor $10^2$, so at $E=10^{20}$ eV it drops from $10^{-49}/cm^2$ down to $10^{-54.2}/cm^2$.
So the reasonable conclusion that the observed values $dJ/dE$ and $F$ are about a few times $10^{-55}$\,cm$^{-2}$. 
The observations exceeds theory by 23 orders  of magnitude.

{The smallness of this result is explained by extremely week annihilation cross-section (\ref{cr-sec}), that is smaller than the 
necessary one (\ref{crsec_Bl}) by 30 orders of magnitude. Our result is not good enough ''only'' by 23 orders of magnitude but not
naively expected 30, since we apply it only to the highest energy tail of UHECR spectrum.
}

{The characteristic time of annihilation, corresponding to (\ref{cr-sec}) is huge:
\be {
\tau_{ann} = n_f /\dot n_f \approx 3\cdot 10^{64}\, \rm{s}},
\label{tau-ann}
\ee
by far exceeding the universe age { $t_U \approx 13.8\ {\text Gyr} \approx 4 \times10^{17}$ sec.}}

{
However, the annihilation can be strongly enhanced due to resonance process of $f\bar f$- transition to scalaron, since the $2 M_f $ is very
close to $M_R$. The resonance effects in dark matter particle annihilation are considered in Refs.~\cite{Griest:1990kh,Gondolo:1990dk}. Eq.~(\ref{dot-n-f})  is valid
for S-wave annihilation and energy independent cross-section. For arbitrary dependence of the cross-section on the center of mass energy
squared $s = (p_f+p_{\bar f})^2$ the average value of $\sigma_{ann} v$ is calculated in Ref.~\cite{Gondolo:1990dk}:
\be
\langle \sigma_{ann} v \rangle = \frac{1}{8 M^4_f T [K_2(M_f/T)]^2} \int_{4M_f^2}^\infty ds\,(s- 4 M^2_f) \sigma_{ann} (s) \sqrt{s} 
K_1\left(\frac{\sqrt{s}}{T}\right) ,
\label{sigma-aver}
\ee
where $T$ is the cosmic plasma temperature, and $K_{(1,2)}$ are the modified Bessel functions. 
Since $x =M_f/T \gg 1$ we can use the asymptotical limit  of  $K_n (x) \approx \sqrt{\pi/(2x)}\, e^{-x}$. Thus
%\be
%\langle \sigma_{ann} v \rangle = \frac{4}{\sqrt{\pi}} \int_0^\infty dz z e^{-z} \sigma_{ann}(s) v %\left( 4M_f^2 (1+ Tz/M_f)\right) .
%\label{sigma-v-2},
%\ee 
\be{
\langle \sigma_{ann} v \rangle = \frac{4}{\sqrt{\pi}} \sqrt{\frac{T}{M_f}}\int_0^\infty dz z e^{-z} \sigma_{ann}(z)  }
\label{sigma-v-2},
\ee
where dimensionless variable $z$ is defined according to $s = 4M^2_f (1+T z /M_f)$. }

{It is proper to remind at this stage that the cross-section of the annihilation is inversely proportional to $v$. For example  
the cross-section $e^+e^-$ - annihilation near threshold is $\sigma (e^+ e^- \rar 2 \gamma) = \pi \alpha^2/(m_e^2 v)$, see e.g.~\cite{ll-4}.
The enhancement of the cross-section due to the long range Coulomb attraction is neglected here. The case of long range interaction of DM particle
may be of interest. }

{The factor in front of Eq.~(\ref{sigma-v-2}) is the thermally averaged relative particle velocity. Indeed, the average velocity of nonrelativistic 
particles with momentum $p$  and velocity $p/M$ in thermal equilibrium is defined as }
\be {
\langle v \rangle = \frac{\int d^3 p (p/M) \exp[-p^2/(2MT)]}{\int d^3 p \exp[-p^2/(2MT)]} = 2 \sqrt{\frac{T}{\pi M}}.}
\label{v-aver} 
\ee
{
This is twice smaller than the coefficient in front of Eq.~(\ref{sigma-v-2}), since the latter is the relative velocity, or better to say, the so called
Moeller velocity.
}

{In our case the cross-section has resonance due to intermediate scalaron state in f anti-f - annihilation, since the mass of the scalaron
is very close to the sum of masses $f$ and $\bar  f$. According to Ref.~\cite{Griest:1990kh} the resonance cross-section has the form:
\be
\sigma_{ann}^{(res)} v = \frac{\alpha ^2 s }{(M_R^2 - s)^2 + M_R^2 \Gamma_R^2},
\label{sigma-res}
\ee
where $M_R = 3\times 10^{13} $ GeV is the scalaron mass and $\Gamma_R $ is its decay width, equal to { (see Eq. \eqref{Gamma-f-old})}
%According to our calculations~\cite{Arbuzova:2021oqa,Arbuzova:2018apk}
$\Gamma_R = M_f^2 M_R/(6M_{Pl}^2)$~\cite{Arbuzova:2021oqa,Arbuzova:2018apk}.
}

Now for thermally averaged resonance cross-section from Eqs.\eqref{sigma-v-2}, \eqref{v-aver} and \eqref{sigma-res} we obtain
\be
\langle \sigma_{res} v \rangle &=&  \int_0^\infty dz z e^{-z} \frac{ \alpha^2 s }{(M_R^2-s)^2 + M_R^2 \Gamma_R^2 } \approx 
\nonumber\\
&\approx& \frac{\alpha^2}{M^2_{ R}} \int_0^\infty \frac{dz z e^{-z}} {(Tz/M_f)^2 + \Gamma_R^2/M^2_{ R}} =
 \frac{\alpha^2}{M^2_{ R}}\int^\infty_0 \frac{ dz z e^{-z}}{\gamma^2 +\eta^2 z^2} ,
\label{sigma-res}
\ee
where { $\gamma^2 =\Gamma_R^2/M_R^2 = 1/36\,(M_f/M_{Pl})^4 \approx 6.7\cdot 10^{-26}$, and 
$\eta^2 = (T/M_f)^2  %=(T/(2.7 \rm{K})^2 (0.0002 eV/(1.5\cdot 10^{13})^2 \,GeV)  
\approx 2.45\cdot 10^{-52} $, where we took $T=T_{CMB} = 2.7K = 2.35 \cdot 10^{-4}$ eV and $M_f = 1.5 \cdot 10^{13}$ GeV. }

{Thus we can neglect the term $\eta^2 z^2$ and conclude that the annihilation cross-section is 26 orders of magnitude higher than
the estimate made above and correspondingly the contribution to the flux of the cosmic rays  might be at the sufficient
level to explain the origin of the ultrahigh energy cosmic rays with $E \gtrsim 10^{20}$ eV.}

{The effect is even stronger in the case of 
%observed flux but it may be helpful in the case of the possible origin of UHECR via 
annihilation of $f\bar f$ in denser regions of the Galaxy, see the next subsection.
}

\subsection{{Flux of UHECR from DM annihilation in the galactic center \label{Gal-DM}}}

In this section we estimate the flux of cosmic rays originating from DM annihilation in the Galactic center , where the local density 
of DM  is much larger than the average cosmological density \cite{Sofue:2020rnl}:
\be
\rho_{GC} = 840 \ \text{GeV/cm}^3.
\label{rho-GC}
\ee
It exceeds the average DM density by 9 orders of magnitude. Since the flux of the cosmic rays from DM annihilation is 
proportional to the square of the DM particle density, smaller objects 
with the number density larger than the average one can create a larger flux of the cosmic rays.  

In Sec.~\ref{s-homDM} the flux of cosmic rays $L$ Eq.~(\ref{F}) from DM annihilation
in the whole galaxy assuming (unrealistic) homogeneous distribution of DM is calculated.
The result obtained should be rescaled as follows. It should be multiplied by the square of the ratio of densities of DM in the galactic center to the
average cosmological density of DM since the rate of annihilation is proportional to $n_f^2$ , Eq.~(\ref{dot-n-f}). Next it is to be multlipied by the
volume of the high density cluster in the Galactic center $4\pi r_{cl}^3/3$
and {divided by the area of the sphere, $4\pi d_{gal}^2$,  at the distance $d_{gal}$ from the galactic center.} 
Thus the following rescaling is to be done:
\be 
L_{gc} = L_{hom} \times \left(\frac{n_{gal}}{\bar n_{dm}}\right)^2 \frac{(r_{cl}^3/(3\,d_{gal}^2)}{R_{max}}, %\left( \frac{d\dot n_{pp}}{dE}) \right),
\label{L-rescale}
\ee
where $L_{hom}$ is determined by Eqs.~(\ref{F}), \eqref{S-of-rho}, (\ref{L}).  The factor
$d \dot n_{pp}/{dE}$,  entering these expressions, is given by Eq.~(\ref{dn-dE}) and $R_{max} =10^{28}$ cm. 

We assume that the size of this
high density clump in the galactic center is about $r_{cl} =10 \rm{pc} \approx 3\times 10^{19}$ cm and its distance
to the Earth is  $d_{gal} =$ 8 kpc = 2.4 $\cdot 10^{22}$ cm.  Thus the flux could be increased by the factor $1.1\times 10^3$.
%{\red EA: I got $1.1\times 10^3$ with factor 3 in \eqref{L-rescale})}
As we have mentioned above the flux created by homogeneously distributed DM is 23 orders of magnitude below the observed value.
So the situation is only slightly better. However, it would be strongly improved if the DM annihilation goes through $R$-resonance, see 
Eq.~(\ref{sigma-res}).

\section{{Flux of cosmic rays from annihilation of DM with realistic distribution in the Galaxy.} \label{real-DM}}

%%%%%

We take the commonly accepted shape of dark matter distribution \cite{Gunn:1972sv}:   
 \be
 {\displaystyle \rho (r)=\rho _{0}\left[1+\left({\frac {r}{r_{c}}}\right)^{2}\right]^{-1}} \equiv \rho_0 q(r) ,
 \label{dm-of-r}
 \ee
 where $\rho _{0}$ denotes the finite central density and $r_{c}$ the core radius. We assume for the sake of estimate $r_{c}=1$~kpc and
 calculate $\rho _{0}$ from the condition that at the position of the Earth  
 at $r = l_\oplus= 8$ kpc  the density of dark matter is $ \rho (l_\oplus)\approx  0.4\, {\rm GeV/cm}^3$~\cite{Salucci:2010qr}. Hence we find: 
 \be  
 \rho_0 = 65 \rho ( l_\oplus) =
 %8\, \text{kpc}) = 
 26 \,{\rm GeV/cm}^3.
 %\rho (8\, \text{kpc})\approx  0.4\, {\rm GeV/cm}^3 
 %0.01 M_{\odot} /\text{pc}^3\approx 4 \cdot 10^{-24} \text{g/cm}^3 \approx 2.4\,  \text{GeV/cm}^3.
 \label{rho-center}
 \ee
 
{This value exceeds the average density of cosmological dark matter, 
$\rho_{DM} = 1$~keV/cm$^3$  by $2.6 \times 10^7$.}

Let us consider the annihilation of DM particles at the point determined by the radius-vector $\vec r$ 
with the spherical coordinates $r, \theta, \phi$ directed from the galactic center. The distance of this point to the Earth is
\be
d_\oplus =\sqrt{ (\vec{l_\oplus} + \vec{r})^2 } =\sqrt{ r^2 + l_\oplus^2 - 2 r\,l_\oplus \cos{\theta}}.
\label{d-plus}
\ee

As it has been done above we recalculate the flux of cosmic rays rescaling Eq.~\eqref{L-rescale} 
by the ratio of density of DM equal to
(\ref{dm-of-r}) and (\ref{rho-center}) to its average cosmological density and by the presented below integral over distribution of DM to $R_{max}$.
Thus we get for the realistic distribution of DM in the Galaxy:
\be 
L_{real} = L_{gc} \left( \frac{26 {\rm GeV}}{840 {\rm GeV}}\right)^2 \frac{{3} d_{gal}^2}{ r_{cl}^3 } J,
\label{L-real}
\ee
where J is the integral over DM distribution:
\be
J= \int \frac{d^3 r q(r)}{d_\oplus ^2} 
= 2\pi  \int \frac{dr r^2 q(r)d\cos \theta}{r^2 + l_\oplus^2 - 2 r  l_\oplus\cos \theta} = 
2\pi \int \frac{dr r q(r)}{ l_\oplus} \ln \frac{l_\oplus+r}{l_\oplus-r}.
\label{J}
\ee
{After change of variables, $r = x l_\oplus$, the integral is reduced to the expression below and is taken numericallly:}
\be
J = 2\pi\, l_\oplus \int_0^1 dx x \left( 1 + 64 x^2\right)^{-1} \ln \frac{1+x}{1-x} = 0.2 \,l_\oplus . 
\label{J-2}
\ee
%{\red EA: Should be without $\rho_0$}

{Thus we obtain $L_{real} = 3\cdot 10^5 L_{gc} $. This is noticeably larger than the flux originating from dense galactic
center and allows for much weaker amplification by the resonance annihilation (\ref{sigma-res}). }
%In the case of annihilation in the Galactic center the result is below the necessary value
%by $10^{20}$, so now are are missing our goal by $ 3\cdot 10^{-15}$.}

\subsection{Annihilation in clusters of DM in the Galaxy \label{DM-ann-clusters}}

 Already in the first papers~\cite{Blasi:2001hr,Blasi:2001rb,Dick:2002kp} on the UHECR production
 via heavy DM particle annihilation 
 it was pointed out that DM particles could form high density clusters, where their annihilation could 
 be strongly enhanced. However, neither the number density of the clumps of DM in the Galaxy,
 nor the density of DM inside them are accurately known.   Some theoretical estimates can be found
in Refs.~\cite{Diemand:2008in,Diemand:2006ik}.

Recent analysis of a possible enhancement 
of the DM annihilation signal from the galactic clumps of DM is performed in Ref.~\cite{Baushev:2015aqa}.
It is argued that the clumps should give the main contribution into this signal, even with very mild 
assumption on their properties. Taking into account theoretical estimates~\cite{Diemand:2008in,Diemand:2006ik},
and our results on the flux of UHECR obtained above in sec.~\ref{Gal-DM} we can safely conclude
that the suggested in {present paper} scenario {might}  explain the origin of the extremely energetic tail
of the cosmic rays.

Quite restrictive limits on dark matter annihilation in galactic cusps is found in Ref. \cite{Delos:2023ipo}, 
however, the limit is valid for masses of DM particles below 120 GeV.  {Thus the annihilation of such particles
does not make any contribution to the highest energy tail of the cosmic rays.}
 
 \section{Conclusion \label{s-conclude}}

 {It is commonly accepted that cosmic rays with energies  below $10^{20}$ eV could be created in
catastrophic astrophysical processes, but above this bound no astrophysical mechanism has been found. 
The attempts to explain more energetic cosmic rays led to consideration of heavy particle decays or annihilation
as possible sources of the  extremely high energy  cosmic rays, EHECR. The masses of such hypothetical particles 
should be equal to the highest energy observed  in the cosmic rays or even exceed it.}

However,  such particles are usually
introduced {\it ad hoc}, just for this purpose with the
properties specially adjusted to do the job, while no fundamental model was suggested.

 Heavy stable particles with energy of the order of the scalaron mass  naturally appear ~\cite{Arbuzova:2021etq} in the
  $R^2$ inflationary model. It is very interesting coincidence that the
  scalaron mass, $M_R$ (\ref{M-R}), is very close to the highest energy observed in the cosmic rays. So it
  is tempting to suggest that the product of the scalaron decays are related to the {extremely } high energy  
 cosmic rays.    
  The only necessary (fine)tuning of the suggested mechanism is the fixing of the
 mass of such candidate for DM carrier quite close to $M_R/2$.
 
 Several possible configurations of DM in the universe have been studied above, 
 such the uniform distribution in the whole
 universe, dense DM clump in the center of the galaxy, and  the realistic DM distribution in the galaxy. 
 Even in the last most
 favourable case the calculated flux was about 15 orders of magnitude below the necessary value. However, there are
 several ways to escape this pessimistic conclusion.  Firstly the taken  value of the annihilation 
 cross-section (\ref{cr-sec} )
{ could be } $10^4$ times larger if DM particle interaction becomes strong at high energies and thus  instead
$\alpha \approx 10^{-2}$ we take $\alpha \sim 1$.

{The clumps of DM considered in Refs.~\cite{Diemand:2008in,Diemand:2006ik} might create up to 
$10^{5}$ amplification of the 
flux. And last, but not the least, resonance annihilation could easily allow for the necessary 10 orders of magnitude amplification of the flux.
One more mechanism of amplification of the annihilation is possible long range interaction between dark matter particles and
antiparticles. In the case of sufficiently strong coupling even bound states  of DM particles could 
be {created} leading to very fast
annihilation. So finally, there are several natural mechanisms which could explain the origin of highest energy cosmic rays.}
 
 It is also interesting to study {the possibility that the heavy}
 DM particles could decay via formation of the virtual black
 holes~\cite{Zeldovich:1976vq,Zeldovich:1977be}. It might open another channel for their 
 identification \cite{Arbuzova:2023dif}.
 
 Probably the most promising sources of the annihilation flux emerge either from the canonically distributed
 dark matter in the galaxy (\ref{dm-of-r}) or from the annihilation inside the clumps of DM \cite{Rubin:2001yw}. These two case 
 create different angular distribution of EHECR and it opens a potential way to distinguish between
 them  hopefully in not so distant future.

\section*{Acknowledgement}
The work of E.V. Arbuzova and A.D. Dolgov was supported by RSF grant 23-42-00066.

\end{document}